\documentclass[10pt,letterpaper]{article}
\usepackage[top=0.85in,left=2.75in,footskip=0.75in]{geometry}

\usepackage{amsmath,amssymb}

\usepackage{changepage}

\usepackage{textcomp,marvosym}

\usepackage{cite}

\usepackage{nameref,hyperref}

\usepackage[right]{lineno}

\usepackage[nopatch=eqnum]{microtype}
\DisableLigatures[f]{encoding = *, family = * }

\usepackage[table]{xcolor}

\usepackage{array}

\newcolumntype{+}{!{\vrule width 2pt}}

\newlength\savedwidth

\raggedright
\usepackage[aboveskip=1pt,labelfont=bf,labelsep=period,justification=raggedright,singlelinecheck=off]{caption}

\makeatletter
\renewcommand{\@biblabel}[1]{\quad#1.}
\makeatother

\usepackage{lastpage,fancyhdr,graphicx}
\usepackage{epstopdf}
\fancyheadoffset[L]{2.25in}
\fancyfootoffset[L]{2.25in}
\begin{document}
\vspace*{0.2in}

% Title must be 250 characters or less.
\begin{flushleft}
{\LARGE
\textbf{\newline{Large-scale moral machine experiment on large language models} % Please use "sentence case" for title and headings (capitalize only the first word in a title (or heading), the first word in a subtitle (or subheading), and any proper nouns).
}}
\newline
% Insert author names, affiliations and corresponding author email (do not include titles, positions, or degrees).
\\
Muhammad Shahrul Zaim bin Ahmad \textsuperscript{1,2},
Kazuhiro Takemoto \textsuperscript{1,3*}
% Name1 Surname\textsuperscript{1,2\Yinyang},
% Name2 Surname\textsuperscript{2\Yinyang},
% Name3 Surname\textsuperscript{2,3\textcurrency},
% Name4 Surname\textsuperscript{2},
% Name5 Surname\textsuperscript{2\ddag},
% Name6 Surname\textsuperscript{2\ddag},
% Name7 Surname\textsuperscript{1,2,3*},
% with the Lorem Ipsum Consortium\textsuperscript{\textpilcrow}
\\
\bigskip
\textbf{1} Department of Bioscience and Bioinformatics, Kyushu Institute of Technology, Iizuka, Fukuoka, Japan
\\
\textbf{2} Faculty of Engineering and Technology, Multimedia University, Melaka, Malaysia
\\
\textbf{3} Data Science and AI Research Center, Kyushu Institute of Technology, Iizuka, Fukuoka, Japan
% \\
\bigskip

% Insert additional author notes using the symbols described below. Insert symbol callouts after author names as necessary.
% 
% Remove or comment out the author notes below if they aren't used.
%
% Primary Equal Contribution Note
% \Yinyang These authors contributed equally to this work.

% Additional Equal Contribution Note
% Also use this double-dagger symbol for special authorship notes, such as senior authorship.
% \ddag These authors also contributed equally to this work.

% Current address notes
% \textcurrency Current Address: Dept/Program/Center, Institution Name, City, State, Country % change symbol to "\textcurrency a" if more than one current address note
% \textcurrency b Insert second current address 
% \textcurrency c Insert third current address

% Deceased author note
% \dag Deceased

% Group/Consortium Author Note
% \textpilcrow Membership list can be found in the Acknowledgments section.

% Use the asterisk to denote corresponding authorship and provide email address in note below.
* takemoto@bio.kyutech.ac.jp

\end{flushleft}
% Please keep the abstract below 300 words
\section*{Abstract}
The rapid advancement of Large Language Models (LLMs) and their potential integration into autonomous driving systems necessitates understanding their moral decision-making capabilities. While our previous study examined four prominent LLMs using the Moral Machine experimental framework, the dynamic landscape of LLM development demands a more comprehensive analysis. Here, we evaluate moral judgments across 52 different LLMs, including multiple versions of proprietary models (GPT, Claude, Gemini) and open-source alternatives (Llama, Gemma), to assess their alignment with human moral preferences in autonomous driving scenarios. Using a conjoint analysis framework, we evaluated how closely LLM responses aligned with human preferences in ethical dilemmas and examined the effects of model size, updates, and architecture. Results showed that proprietary models and open-source models exceeding 10 billion parameters demonstrated relatively close alignment with human judgments, with a significant negative correlation between model size and distance from human judgments in open-source models. However, model updates did not consistently improve alignment with human preferences, and many LLMs showed excessive emphasis on specific ethical principles. These findings suggest that while increasing model size may naturally lead to more human-like moral judgments, practical implementation in autonomous driving systems requires careful consideration of the trade-off between judgment quality and computational efficiency. Our comprehensive analysis provides crucial insights for the ethical design of autonomous systems and highlights the importance of considering cultural contexts in AI moral decision-making.

% Please keep the Author Summary between 150 and 200 words
% Use first person. PLOS ONE authors please skip this step. 
% Author Summary not valid for PLOS ONE submissions.   
% \section*{Author summary}
% Lorem ipsum dolor sit amet, consectetur adipiscing elit. Curabitur eget porta erat. Morbi consectetur est vel gravida pretium. Suspendisse ut dui eu ante cursus gravida non sed sem. Nullam sapien tellus, commodo id velit id, eleifend volutpat quam. Phasellus mauris velit, dapibus finibus elementum vel, pulvinar non tellus. Nunc pellentesque pretium diam, quis maximus dolor faucibus id. Nunc convallis sodales ante, ut ullamcorper est egestas vitae. Nam sit amet enim ultrices, ultrices elit pulvinar, volutpat risus.

%\linenumbers

% Use "Eq" instead of "Equation" for equation citations.
\section*{Introduction}
The rapid advancement of Large Language Models (LLMs) has inaugurated a new era in artificial intelligence (AI), demonstrating unprecedented capabilities across diverse domains \cite{fraiwan2023review,sallam2023chatgpt,ray2023chatgpt,chowdhury2024harnessing}.
These models, exemplified by ChatGPT \cite{chatgpt,achiam2023gpt} (developed by OpenAI) and its counterparts, exhibit remarkable proficiency in natural language processing, content generation, complex problem-solving, and decision-making.
As these models advance from laboratory experiments to real-world applications, understanding their decision-making processes, particularly in ethical contexts, becomes crucial for ensuring their responsible deployment. This has become a subject of intense scientific investigation \cite{bostrom2018ethics,nath2020problem,hagendorff2020ethics}.

Autonomous driving represents a field where LLM application necessitates particularly careful evaluation \cite{bonnefon2016social,awad2018moral,faulhaber2019human,gill2021ethical}.
When autonomous vehicles make decisions on public roads, their choices directly affect human safety and well-being.
The automotive industry's investigation of LLM integration for enhancing autonomous vehicles' capabilities includes various aspects: analyzing complex road conditions, interpreting traffic situations, planning appropriate responses, and importantly, making ethical decisions in challenging scenarios \cite{chen2023feedback,gao2023chat,du2023chat,lei2023chatgpt,yang2023llm4drive,cui2024survey,zhou2024vision}.
This integration spans critical functions from environmental comprehension to decision making, highlighting the need for thorough understanding of these models' capabilities and limitations.

The Moral Machine experiment \cite{awad2018moral} provides a systematic framework for evaluating ethical decision-making in autonomous systems.
This study presented participants with scenarios involving unavoidable accidents to assess how autonomous vehicles should behave in morally challenging situations.
The experiment revealed clear patterns in human moral preferences, such as prioritizing human lives over animals and favoring the preservation of a greater number of lives.
While acknowledging its limitations including methodological constraints in scenario design, challenges in cultural representation, and gaps between theoretical choices and practical implementation \cite{dewitt2019moral,bigman2020life,etienne2021dark,furey2021s,lacroix2022moral,schuessler2024probability}, this work has stimulated extensive research into the ethical implications of AI in autonomous systems \cite{winfield2019machine,etienne2022ai,meyer2022baby,atakishiyev2024explainable}.
These foundational insights, despite their limitations which we explore in detail in our discussion, have provided crucial frameworks for examining ethical decision making in autonomous systems.

While some studies have explored AI responses to standard ethical dilemmas \cite{krugel2023chatgpt,jin2024multilingual}, such as the classic trolley problem \cite{bruers2014review}, the intricate scenarios presented in the Moral Machine experiment offer a more comprehensive exploration of AI moral preferences.
Building on this approach, our previous study \cite{takemoto2024moral} applied the Moral Machine framework to a limited set of LLMs, specifically focusing on ChatGPT (including both GPT-3.5 and GPT-4 versions), Llama 2, and PaLM 2.
This investigation examined the moral judgments made by these AI systems in complex autonomous driving scenarios.
The results indicated that LLMs were capable of producing ethical judgments that often aligned with human preferences in many aspects.
However, we also observed some notable discrepancies between LLM outputs and human choices, particularly in scenarios involving multiple ethical considerations

Despite these initial findings, the rapid evolution of LLM technology necessitates a more comprehensive analysis.
This dynamic landscape is evidenced by the frequent release and update of both proprietary models like ChatGPT, Claude \cite{anthropic2024claude} (Anthropic), and Gemini \cite{team2023gemini} (Google), as well as open-source alternatives such as Llama 3 \cite{dubey2024llama} (Meta) and Gemma \cite{team2024gemma} (Google).
The emergence of these diverse models underscores the importance of examining the consistency and evolution of moral judgments across different LLM versions, architectures, and development approaches.

This study aims to conduct a more extensive examination of moral judgment trends in the latest LLM models and investigate the effects of model updates on these judgments.
By comparing a wider array of LLMs, we seek to provide insights into the approaches different LLMs take to moral dilemmas, how these approaches may change over time, and the potential implications for AI system deployment in critical decision-making roles, particularly in autonomous driving.
This study will contribute to a more nuanced understanding of AI outputs in high-stakes scenarios, informing the development and deployment of AI systems in complex, real-world contexts.

\section*{Materials and methods}
\subsection*{Moral Machine scenario generation}
To comprehensively evaluate how different LLMs approach moral decision-making in autonomous driving scenarios and assess the effects of model updates on these judgments, we generated a systematic set of test scenarios following our previous methodology.
These scenarios presented dilemmas involving an autonomous vehicle facing an unavoidable accident, requiring a choice between two outcomes (``Case 1'' and ``Case 2'') with varying ethical implications.

The scenarios were designed through a process of constrained randomization, exploring six primary dimensions:
species (humans vs. pets),
social value (higher vs. lower perceived social status),
gender (female vs. male),
age (younger vs. older),
fitness (physically favored vs. less fit individuals), and
utilitarianism (smaller vs. larger group).
Additionally, three secondary dimensions were incorporated:
interventionism (swerving vs. continuing straight),
relationship to the autonomous vehicle (passengers vs. pedestrians), and
legal considerations (e.g., adherence to traffic signals).
The scenarios were carefully constructed to balance complexity and clarity, ensuring that they effectively probed the ethical reasoning capabilities of the LLMs under study.

\subsection*{Large language models}
To investigate how moral judgment patterns vary across different model architectures and evolve through updates, we expanded upon our previous study \cite{takemoto2024moral} by incorporating a comprehensive set of LLMs, including the latest versions available.
Our analysis encompasses both base language models and models specifically optimized for dialogue interactions.
While these models may differ in their training approaches, they share fundamental capabilities in processing and responding to complex queries, including moral scenarios. 
These LLMs were utilized to generate responses to moral machine scenarios.

We initially incorporated the results from \cite{takemoto2024moral}, which included responses from three proprietary models: GPT-3.5 (gpt-3.5-turbo-0613), GPT-4 (gpt-4-0613), and PaLM 2, as well as one open source model: Llama 2 (7B).
Additional models were selected based on their prominence in the field, technological advancements, and public recognition.
This selection aimed to provide a representative sample of current LLM technology while exploring variations in ethical reasoning across different model types.

Among the proprietary models, we evaluated multiple versions of OpenAI models, including the snapshot versions of GPT-3.5 \cite{chatgpt} and GPT-4 \cite{achiam2023gpt} from March 2023 to April 2024, as well as the newer GPT-4o models \cite{GPT4o} (May and August 2024 versions), GPT-4o-mini, o1 and o1-mini \cite{OpenAIo1}.
Google DeepMind's Gemini models were also assessed, including Gemini 1.0 Pro \cite{team2023gemini} and multiple snapshot versions of Gemini 1.5 Pro and Gemini 1.5 Flash \cite{reid2024gemini}.
From Anthropic, we evaluated the latest versions of their Claude models, including Claude 3 Haiku, Sonnet, Opus \cite{anthropic2024claude}, and Claude 3.5 Sonnet \cite{anthropic2024claude3.5}.

In the open-source domain, we included Meta's Llama family \cite{dubey2024llama}, comprising Llama 3 and Llama 3.1 in 8B and 70B versions, Llama 3.2 in 1B and 3B versions, and Llama 3.3 in 70B version.
A comprehensive evaluation of Google's Gemma model family \cite{team2024gemma,team2024gemma2} was conducted, encompassing multiple versions and scales from 2B-it to 27B-it, including the derivative DataGemma RIG 27B-it model \cite{radhakrishnan2024knowing}.
Additional open-source models evaluated in this study included Large Model Systems Organization's Vicuna v1.5 (7B and 13B) \cite{vicuna2023}, Mistral (7B Instruct v0.2) \cite{jiang2023mistral}, Mistral-NeMo \cite{MistralNeMo}, CohereForAI's Command R+ \cite{CommandRplus2024}, and Microsoft's Phi-3.5 \cite{abdin2024phi} (both MoE-instruct and mini-instruct versions).

For a comprehensive list of all models and their specific versions used in this study, please refer to our GitHub repository (https://github.com/kztakemoto/mmllm).

All models were used with their default parameter settings.
The input prompts were consistent with those used in \cite{takemoto2024moral}, requiring each model to select either ``Case 1'' or ``Case 2'' for each scenario.
The complete code for prompt generation, analysis, and resulting data is available in our GitHub repository.

While most models were evaluated using 50,000 scenarios, following our previous study \cite{takemoto2024moral}, constraints related to application programming interface (API) usage costs and computational time necessitated the use of fewer scenarios for certain models.
The majority of the models were assessed using the full set of 50,000 scenarios. However, a substantial group of models, including the GPT-4 and GPT-4o series, Claude 3 Opus and Claude 3.5 family, all Gemini family, Llama 3 and subsequent versions, OpenAI o1 mini, Mistral-NeMo, Gemma-2 family, DataGemma, Command R+, and Phi-3.5 family, were evaluated using a reduced set of 10,000 scenarios.
The OpenAI o1 model, facing even more significant API cost limitations, was evaluated using a further reduced set of 5,000 scenarios.

\subsection*{Data analysis}
To systematically evaluate how different LLMs align with human moral preferences and how this alignment changes across model versions and architectures, we analyzed the LLM responses using the conjoint analysis framework as described in \cite{hainmueller2014causal} and implemented in the original study \cite{awad2018moral} on the Moral Machine experiment and our previous study \cite{takemoto2024moral}.
This approach allows for a robust, non-parametric identification of causal effects without specific modeling assumptions.

The analysis process involved pre-processing LLM responses and dummy variable coding for scenario attributes, which correspond to the primary and secondary dimensions described in the scenario generation.
We then calculated the Average Marginal Component Effect (AMCE) for each attribute. The AMCE quantifies an attribute's influence on the LLM's ethical decisions, enabling systematic comparison of moral preferences across different LLM models and versions.

To compare the moral preference patterns between LLMs and humans, we performed comparative analyses using the calculated AMCE values across all nine attributes. The human preference data were obtained from the original Moral Machine experiment \cite{awad2018moral}.
We quantified the alignment between LLM and human preferences by computing the Euclidean distances between their respective AMCE values. Furthermore, to better understand the relationship patterns among different models and human preferences, we employed principal component analysis (PCA) followed by cluster analysis on the AMCE values.

All statistical analyses, including AMCE calculations, distance measurements, PCA, and subsequent analyses were performed using R statistical software (version 4.4.1).

% Results and Discussion can be combined.
\section*{Results}
\subsection*{Moral judgments of large language models}
Moral judgments of large language models
To comprehensively evaluate how different LLMs approach moral decision-making in autonomous driving scenarios, we conducted a systematic analysis of their responses across multiple ethical dimensions.
Our analysis framework focused on quantifying and comparing the moral preferences of various LLM families with human judgments.

\begin{figure}[!h]
\begin{adjustwidth}{-2.25in}{0in} 
\begin{center}
\includegraphics[width=170mm]{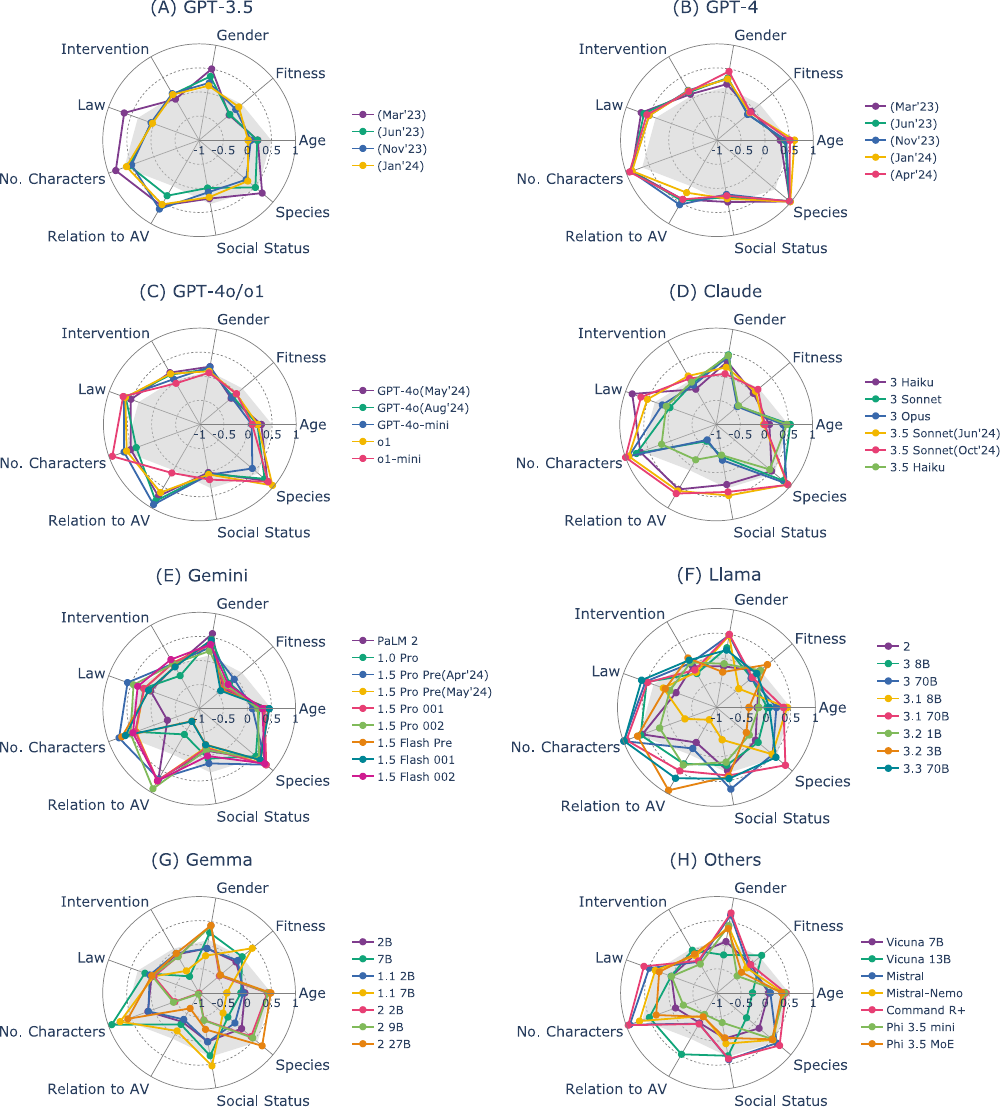} % Here is how to import EPS art
\end{center}
\caption{{\bf Radar plots of moral preferences across different LLM families.}
AMCE values indicate preferences: Species ($+$: humans, $-$: pets), Social Value ($+$: high status, $-$: low status), Relation to AV ($+$: pedestrians, -: passengers), Number ($+$: more, $-$: fewer), Law ($+$: lawful, $-$: unlawful), Intervention ($+$: inaction, $-$: action), Gender ($+$: female, $-$: male), Fitness ($+$: fit, $-$: unfit/obese), Age ($+$: young, $-$: elderly).
Gray-filled areas represent human preferences.
Each subplot represents a different model family: (A) GPT-3.5, (B) GPT-4, (C) GPT-4o/o1, (D) Claude, (E) Gemini, (F) Llama, (G) Gemma, and (H) Other models.}
\label{fig:llm_preference_radar_chart}
\end{adjustwidth}
\end{figure}

Fig \ref{fig:llm_preference_radar_chart} presents radar charts characterizing the moral judgment tendencies of various LLMs across nine preferences quantified by Average Marginal Component Effect (AMCE) values (see \nameref{S1_Table} for exact numerical data).
These charts enable direct comparison of preferences across multiple versions and model sizes of major LLM families, including GPT-3.5, GPT-4, GPT-4o/o1, Claude, Gemini, Llama, Gemma, and other LLMs, against established human preferences (see \nameref{S1_Fig} for detailed differences between each model and human values).

Our initial analysis revealed three key patterns in LLM moral judgments.
First, many LLMs demonstrated strong alignment with human preferences in fundamental moral decisions, showing large positive values ($>0.5$) for No. Characters and Species dimensions.
This indicates consistent tendencies in prioritizing the preservation of more lives and favoring humans over animals, suggesting that LLMs can capture basic human moral intuitions.
Second, we observed significant variations in judgment patterns across different model families, with some models exhibiting inverse priorities compared to human preferences.
Third, certain models showed disproportionately strong preferences in the same direction as humans (differences in AMCE values approximately $>0.3$ compared to human values), indicating potential overemphasis of specific moral principles.

These patterns provide important insights into both the capabilities and limitations of LLMs in moral reasoning tasks. Below, we elaborate on these findings for each model family, examining how different architectures influence moral judgments, and analyzing the effects of model updates on alignment with human preferences.

\subsubsection*{GPT-3.5 family}
Analysis of the GPT-3.5 family (Fig \ref{fig:llm_preference_radar_chart}A) revealed significant temporal evolution in moral judgments across model updates.
While models from June 2023 to January 2024 exhibited similar patterns, the March 2023 version showed distinctive tendencies, particularly in several key ethical dimensions.

In the Law dimension, the March version showed strong positive values favoring law-abiding individuals (AMCE = 0.7), but subsequent versions shifted toward more neutral judgments (AMCE $\approx$ 0).
Similar temporal changes were observed in Gender-related judgments, where initial strong female preference (AMCE = 0.5 in March 2023) gradually reduced toward human judgment levels (AMCE = 0.2 in January 2024, human AMCE = 0.1).
Fitness-related judgments also evolved from initially contradicting human preferences (AMCE $=-0.2$ in March 2023) to showing better alignment in later versions (AMCE = 0.1 in January 2024).

However, we observed persistent deviations from human preferences in certain dimensions.
The latest versions maintained near-neutral age-related judgments (AMCE $\approx 0$) despite human preference for younger individuals (AMCE = 0.5), and showed consistently stronger preferences for pedestrians over passengers (AMCE = 0.6 vs. human 0.1).
Species-related judgments demonstrated an interesting pattern where initial alignment with human values in earlier versions (AMCE = 0.7) weakened in later updates (AMCE = 0.2) while maintaining the same directional preference.

\subsubsection*{GPT-4 family}
Analysis of the GPT-4 family (Fig \ref{fig:llm_preference_radar_chart}B) revealed remarkable consistency in moral judgments across different versions from March 2023 to January 2024.
In contrast to GPT-3.5, model updates showed minimal impact on judgment patterns, with all versions maintaining similar preference strengths across ethical dimensions.

The GPT-4 family demonstrated strong alignment with human judgments in direction but often exceeded human preference magnitudes.
Most notably, Species and No. Characters preferences showed near-deterministic values (AMCE $>0.9$) compared to human values (AMCE = 0.6), indicating substantially stronger preferences for saving humans over animals and larger groups over smaller ones.
Similar amplification appeared in Gender (AMCE = 0.4 vs. human 0.1), Law (AMCE = 0.6 vs. human 0.3), and Relation to AV (AMCE = 0.5 vs. human 0.1) categories.

The only notable divergence appeared in the Fitness category, where GPT-4 models showed a slight preference for overweight individuals (AMCE $= -0.1$) contrary to human preference for physically fit individuals (AMCE = 0.2).
This deviation, however, was minimal compared to the strong alignment observed in other dimensions.

\subsubsection*{GPT-4o/o1 family}
Analysis of the GPT-4o family (Fig 1\ref{fig:llm_preference_radar_chart}) revealed a pattern of consistent moral judgments across models, with two notable exceptions: GPT-4o-mini and o1-mini showed distinctive characteristics.
The family exhibited closest alignment with human preferences in Gender (AMCE = 0.1) and No. Characters (AMCE = 0.7) categories, matching human values within $\pm 0.1$ AMCE.

Most GPT-4o models showed stronger ethical preferences than both humans and the GPT-4 family in several dimensions.
In Relation to AV, all models except GPT-4o-mini demonstrated extreme pedestrian preference (AMCE $> 0.5$ vs. human 0.1). Similarly, they showed amplified preferences for law-abiding individuals (AMCE = 0.8 vs. human 0.4) and humans over pets (AMCE = 0.9 vs. human 0.6).
However, these models showed weaker age-related preferences (AMCE = 0.2 vs. human 0.5) and maintained slight preferences for overweight individuals in Fitness judgments (AMCE $= -0.1$).

The o1 variants showed interesting divergences: while the standard o1 followed GPT-4o patterns, o1-mini demonstrated unique characteristics, including near-neutral Relation to AV preferences (AMCE = 0.1) but extremely strong preferences for larger groups (AMCE $\approx$ 1.0).
This suggests that model size significantly influences moral judgment patterns within this family.

\subsubsection*{Claude family}
Analysis of the Claude family (Fig \ref{fig:llm_preference_radar_chart}D) revealed a clear bifurcation in moral judgment patterns, forming two distinct groups with contrasting ethical preferences.
The first group (Claude 3 Sonnet, Opus, and 3.5 Haiku) and second group (Claude 3 Haiku and Claude 3.5 Sonnet) showed markedly different alignment with human preferences.

The first group demonstrated significant divergence from human judgments in several dimensions.
They showed inverse preferences in Relation to AV (AMCE $= -0.4$), Fitness (AMCE $\approx -0.5$), and Social Status (AMCE $= -0.3$) categories, contrasting with human values (AMCE = 0.1, 0.2, and 0.3 respectively).
While they aligned with humans in Gender and Species preferences, their magnitude was notably stronger in both dimensions (Gender: AMCE = 0.7 vs. human 0.3; Species: AMCE = 0.9 vs. human 0.6).

The second group showed different patterns: Claude 3 Haiku and Claude 3.5 Sonnet exhibited excessive preferences in Relation to AV (AMCE = 0.5) and Law (AMCE = 0.7) categories compared to humans (AMCE = 0.1 and 0.4, respectively).
Claude 3.5 Sonnet particularly showed strong preferences in Species and No. Characters dimensions (AMCE $> 0.8$).

Model updates showed inconsistent effects: while the Sonnet 3 to 3.5 update improved human alignment, the Haiku 3 to 3.5 update increased divergence.
Subsequent updates within Claude 3.5 Sonnet showed minimal changes in judgment patterns.

\subsubsection*{Gemini family}
Analysis of the Gemini family (Fig \ref{fig:llm_preference_radar_chart}E) revealed significant evolution in moral judgments across model updates, particularly marked by changes in the Relation to AV dimension.
We observed a clear transition from Gemini 1.0 Pro to the 1.5 series, with distinct changes in ethical preferences.

The most notable shift occurred in the Relation to AV category, where Gemini 1.0 Pro initially showed preference for passengers (AMCE $= -0.4$), while the Gemini 1.5 Pro series shifted to strongly favor pedestrians (AMCE = 0.7 vs. human 0.1). The Gemini 1.5 Flash family showed a similar evolution: Flash Preview and 001 favored passengers (AMCE $= -0.7$), but Flash 002 shifted to strong pedestrian preference (AMCE = 0.7).

Gender and Species preferences also showed distinctive patterns. Only Gemini 1.5 Pro 002 aligned closely with human gender preferences (AMCE = 0.2 vs. human 0.1), while earlier versions showed stronger female preference (AMCE $> 0.4$).
All versions showed strong preferences than humans for prioritizing humans over pets (AMCE $> 0.7$) and consistently showed inverse fitness-related preferences (AMCE $= -0.3$ vs. human 0.2).

Notably, compared to Google's previous PaLM 2 model, which contradicted human preferences in group size decisions (AMCE = $-0.3$ vs. human 0.7), all Gemini versions maintained consistent alignment with human preferences in preserving larger groups (AMCE = 0.4--0.8).

\subsubsection*{Llama family}
Analysis of the Llama family (Fig \ref{fig:llm_preference_radar_chart}F) revealed that model size significantly influenced moral judgment patterns, with larger models showing more consistent alignment with human preferences.
Most models demonstrated human-like preferences in the No. Characters category (AMCE = 0.2--0.9), with Llama 3.1 8B being the notable exception (AMCE $= -0.3$).

Ethical preferences varied substantially across different model sizes and versions. In gender-related decisions, smaller models (Llama 2 7B, 3.2 1B, 3.2 3B) showed male preference (AMCE $<-0.1$) contrary to human judgments (AMCE = 0.1), while larger models aligned with human preferences.
Species-based decisions showed similar size-dependent patterns: 70B models (Llama 3, 3.1, and 3.3) demonstrated close alignment with human judgments (AMCE = 0.5--0.6), while smaller models (e.g., Llama 3.2 1B and 3B) showed inverse preferences (AMCE $<-0.1$).

The 70B models showed relatively small variations in judgments between versions and demonstrated closer alignment with human preferences.
For example, these models maintained consistent preferences in law compliance (AMCE = 0.5) which happened to align with human values (0.4), while other models, excluding 3 8B model, exhibited neutral preferences.
This observation suggests that model size may influence the consistency of judgment patterns.

\subsubsection*{Gemma family}
Analysis of the Gemma family (Fig \ref{fig:llm_preference_radar_chart}G) revealed a mix of consistent patterns and model-dependent variations in moral judgments.
Across all model variations, we observed consistent trends in law compliance (AMCE $\approx$ 0) and Relation to AV preferences (AMCE $<0$), with the latter diverging from human judgments (human AMCE = 0.1).

Judgment patterns varied significantly across different model sizes and versions.
In the No. Characters category, smaller Gemma 2 models (2B, 9B) showed inverse preferences (AMCE $= -0.4$) compared to human values (AMCE = 0.7), while larger models aligned with human preferences.
The Species category showed similar variation: the Gemma 2 series demonstrated human-aligned preferences (AMCE = 0.6), while other versions showed weak or neutral species preferences (AMCE $< 0$).
Social Status preferences varied notably, with only Gemma 7B and 1.1 7B matching human preferences (AMCE = 0.3), while other versions showed neutral or inverse preferences.

The variation in judgment patterns across different model sizes suggests that parameter count significantly influences moral reasoning capabilities, although this relationship appears more complex than in other model families.

\subsubsection*{Other large language models}
Analysis of other LLMs (Fig \ref{fig:llm_preference_radar_chart}H) revealed distinct patterns of alignment and divergence from human moral preferences, with notable variations across model architectures.
In the Gender category, most models except Vicuna showed substantially stronger female preference (AMCE $> 0.4$) compared to human values (AMCE = 0.1).

Key differences emerged in utilitarian judgments: while Mistral and Command R+ aligned with humans in favoring larger groups, their preferences were markedly stronger (AMCE = 0.9 vs. human 0.6).
Characteristic divergences appeared in the Relation to AV and Fitness categories, where most models except Vicuna 13B showed inverse preferences (Relation to AV: AMCE $<0$ vs. human 0.1; Fitness: AMCE $<0$ vs. human 0.2).

Vicuna models demonstrated unique patterns. While Vicuna 7B showed neutral judgments (AMCE $\approx$ 0) across multiple categories (Gender, Age, Species), Vicuna 13B exhibited distinctive preferences contrary to humans (Gender: AMCE = $-0.2$ vs. human 0.1; Species: AMCE = $-0.2$ vs. human 0.6).
Similarly, Phi 3.5 models showed size-dependent variations, with the mini version demonstrating notably different preferences than the MoE version.
These results suggest potential influences of model size on ethical judgments.

\subsection*{Quantitative comparison with human moral judgments}
\begin{figure}[!h]
\begin{adjustwidth}{-2.25in}{0in} 
\begin{flushright}
\includegraphics[width=170mm]{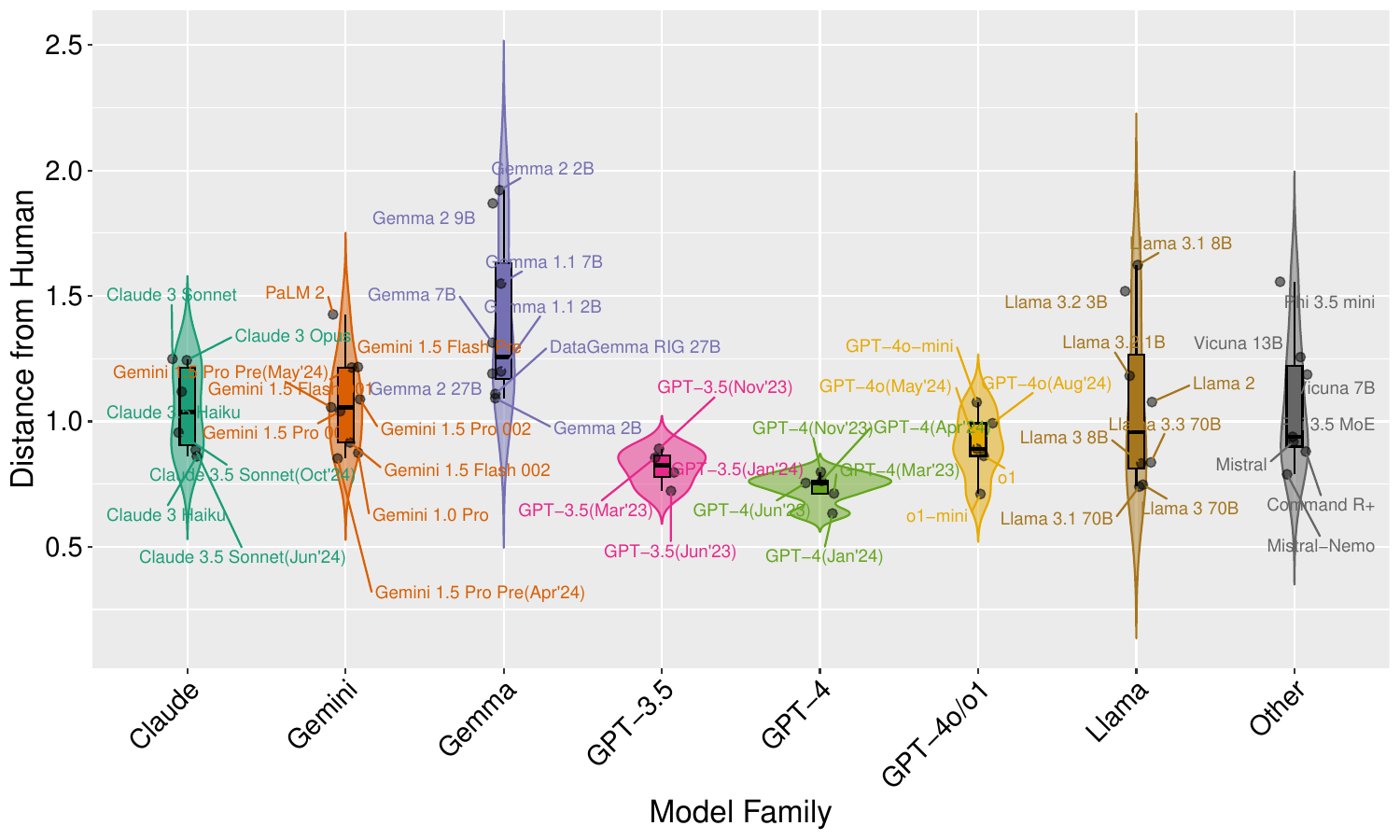} % Here is how to import EPS art
\end{flushright}
\end{adjustwidth}
\caption{{\bf Distances between LLMs and human moral judgments across model families.}
Violin plots show the distribution of distances from human judgments for each model family, with individual models represented as points. Different colors indicate different model families. Model names are labeled.}
\label{fig:distance_wrt_modeltype}
\end{figure}

To quantitatively evaluate the similarity between LLM and human moral judgments, we compared the Euclidean distances of AMCE values across nine categories for each model family (Fig \ref{fig:distance_wrt_modeltype}; see \nameref{S2_Table} for exact numerical data).
This analysis revealed differences between model families and changes in judgment distances from humans through model evolution.

Proprietary models generally showed closer alignment with human judgments than open-source alternatives.
The GPT-4 family achieved the closest alignment (minimum distance = 0.6 with January 2024 version), followed by GPT-3.5 (distance = 0.8--0.9).
However, the newer GPT-4o/o1 family showed slightly increased distances (0.9--1.0), with o1-mini being an exception (distance = 0.7).

Model updates showed varying effects across different families.
In the Claude family, while Claude 3 Haiku and Claude 3.5 Sonnet demonstrated relatively close alignment with human judgments (distances = 0.9--1.0), Claude 3 Sonnet and Opus showed notably larger distances (1.2).
The Gemini family similarly exhibited significant version-dependent variations, with Gemini 1.5 Preview (April 2024) showing the minimum distance of 0.9, while subsequent versions sometimes demonstrated larger distances of 1.1.
Additionally, 1.5 Flash models showed greater distances compared to 1.5 Pro models.

Open-source model performance strongly correlated with model size.
The Llama family showed clear size-dependent improvement, with 70B models achieving closer alignment (distance = 0.7--0.8) compared to smaller 1--8B models (distances = 1.2--1.6).
Conversely, recent Gemma 2 models showed larger distances (2B/9B: distance = 2.0) despite being newer, though performance improved with size.
Among other LLMs, Mistral-Nemo and Command R+ showed relatively good alignment (distances = 0.8--0.9), while Vicuna and Phi 3.5 demonstrated larger divergence (distances = 1.0--1.6).

\subsection*{Factors influencing judgment distance}
\subsubsection*{Proprietary vs. open-source models}
To identify factors influencing the distance from human judgments, we first investigated whether the proprietary or open-source nature of models influenced their alignment with human moral judgments. This analysis was motivated by previous research suggesting general performance advantages of proprietary models over open-source alternatives \cite{zhang2024closing}.

\begin{figure}[!h]
%\begin{adjustwidth}{-2.25in}{0in} 
\begin{center}
\includegraphics[width=120mm]{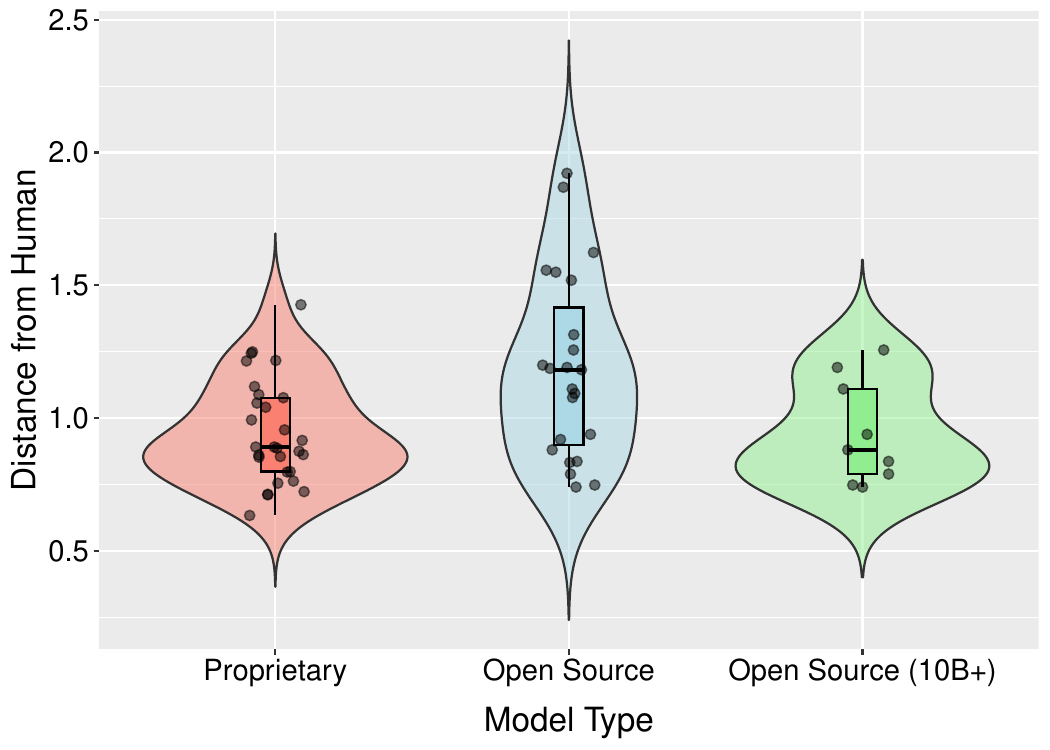} % Here is how to import EPS art
\end{center}
%\end{adjustwidth}
\caption{{\bf Comparison of moral judgment distances between proprietary and open-source models}
Violin plots with embedded box plots compare the distribution of distances from human judgments across three model categories: proprietary models, all open-source models, and large open-source models with parameters exceeding 10B. Individual models are represented as points.}
\label{fig:distance_wrt_opensource}
\end{figure}

Our analysis classified models into three distinct groups (Fig \ref{fig:distance_wrt_opensource}): proprietary models (e.g., GPT-4, Claude, Gemini), general open-source models, and large open-source models ($>$10B parameters). Statistical comparison revealed significant differences between proprietary and general open-source models, with proprietary models showing closer alignment to human judgments (median distance: 0.9 vs. 1.2, Wilcoxon test $p = 0.011$).

However, when comparing proprietary models with large open-source models specifically, this difference disappeared. Large open-source models ($>$10B parameters) achieved similar alignment with human judgments (median distance = 0.9) as proprietary models, showing no statistically significant difference (Wilcoxon test $p = 0.92$).
Notably, models like Llama-70B achieved distances (0.7--0.8) comparable to or better than some proprietary models.

These findings suggest that model size, rather than the proprietary/open-source distinction, is the primary factor influencing alignment with human moral judgments. This observation has important implications for the development of ethical AI systems, suggesting that large open-source models can potentially match the moral reasoning capabilities of proprietary alternatives.

\subsubsection*{Impact of model size}
\begin{figure}[!h]
%\begin{adjustwidth}{-2.25in}{0in} 
\begin{center}
\includegraphics[width=120mm]{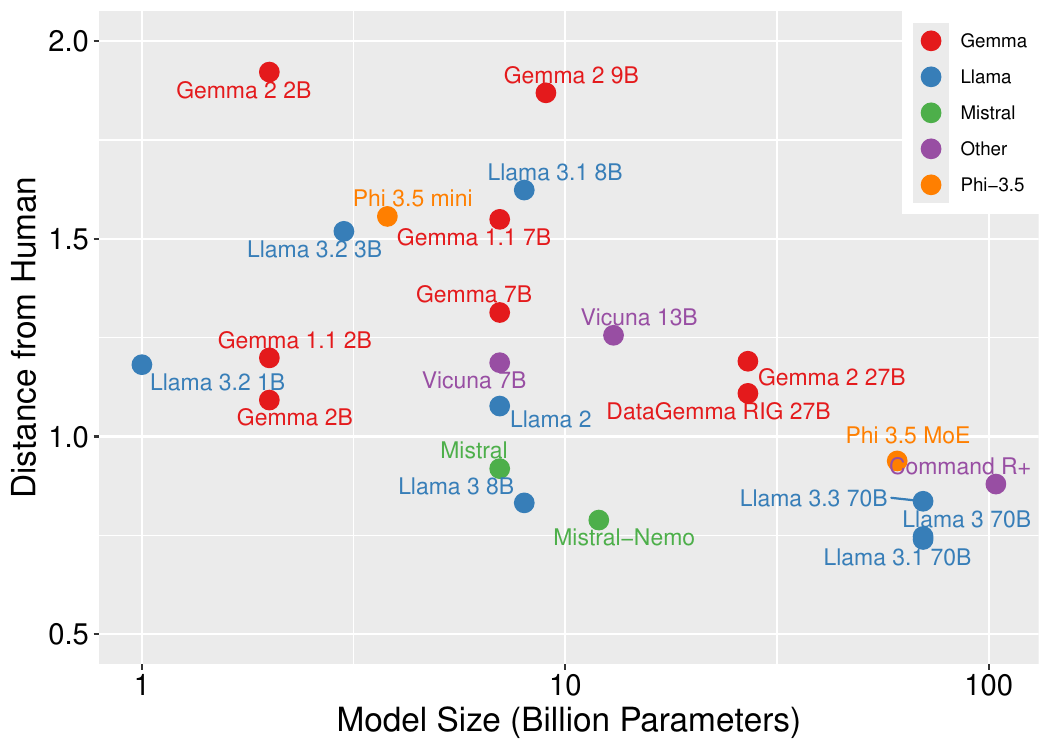} % Here is how to import EPS art
\end{center}
%\end{adjustwidth}
\caption{{\bf Relationship between model size and distance from human judgments in open-source models.}
Different model families are represented by different colors. Model names are labeled. Horizontal axis (model size in billion parameters) are represented in logarithmic scale.}
\label{fig:modelparam_vs_distance}
\end{figure}

We analyzed the relationship between model size and alignment with human moral judgments, focusing on open-source models where parameter counts are publicly available (Fig \ref{fig:modelparam_vs_distance}). Statistical analysis revealed a significant negative correlation between model size and distance from human judgments (Spearman's rank correlation coefficient $\rho = -0.54$, $p = 0.0076$), though this relationship showed notable complexities.

The Llama family demonstrated the clearest size-dependent improvements in alignment.
Large (70B) models achieved substantially better alignment (distance = 0.7--0.8) compared to smaller models (1-3B parameters, distances = 1.2--1.5).
A similar pattern emerged in the Gemma family, where the 27B model showed improved alignment (distance = 1.1) compared to smaller 2--9B models (distances = 1.2--1.9).
These improvements suggest that increased model capacity enables more nuanced moral judgments.

However, the relationship between size and alignment showed important nuances within model families.
Within the Llama family, the 3.2 series showed unexpected variation, with the 1B model achieving better alignment (distance = 1.2) than the larger 3B model (distance = 1.5). Similar inconsistencies appeared among 8B models, where version 3 demonstrated notably better alignment (distance = 0.8) than the subsequent version 3.1 (distance = 1.6). The Gemma family showed analogous patterns, with some larger models demonstrating increased distances despite their greater parameter counts.

These variations suggest that while model size significantly influences moral judgment alignment, other factors such as architecture design and training methodology play crucial roles. The moderate correlation coefficient ($\rho = -0.54$) quantitatively supports this observation.

While proprietary models were excluded from this analysis due to undisclosed parameter counts, their generally strong alignment with human judgments (median distance = 0.9) is consistent with the observed trend that larger models tend to demonstrate better alignment with human moral preferences.

\subsubsection*{Impact of model updates}
Analysis of model updates revealed that newer versions did not consistently improve alignment with human moral judgments, suggesting a complex relationship between model evolution and ethical reasoning capabilities. We examined this relationship across both open-source and proprietary model families, finding significant variations in the impact of updates.

In open-source models, updates often showed unpredictable effects on moral judgment alignment.
Within the Gemma family, successive updates led to increasing divergence from human judgments: the 2B model showed progressive increase in distances (Gemma 2B: 1.1, Gemma 1.1 2B: 1.2, Gemma 2 2B: 2.0).
The Llama family exhibited mixed effects: in 7-8B models, while the update from Llama 2 to 3 improved alignment (distance decreased from 1.1 to 0.8), the subsequent update to 3.1 significantly increased divergence (distance = 1.6).
In the 70B series, while Llama 3 and 3.1 maintained strong alignment (distance = 0.7), the subsequent Llama 3.3 showed a slight increase in divergence (distance = 0.8).
These results indicate that even for larger models, updates do not consistently improve alignment with human moral preferences.

\begin{figure}[!h]
%\begin{adjustwidth}{-2.25in}{0in} 
\begin{center}
\includegraphics[width=120mm]{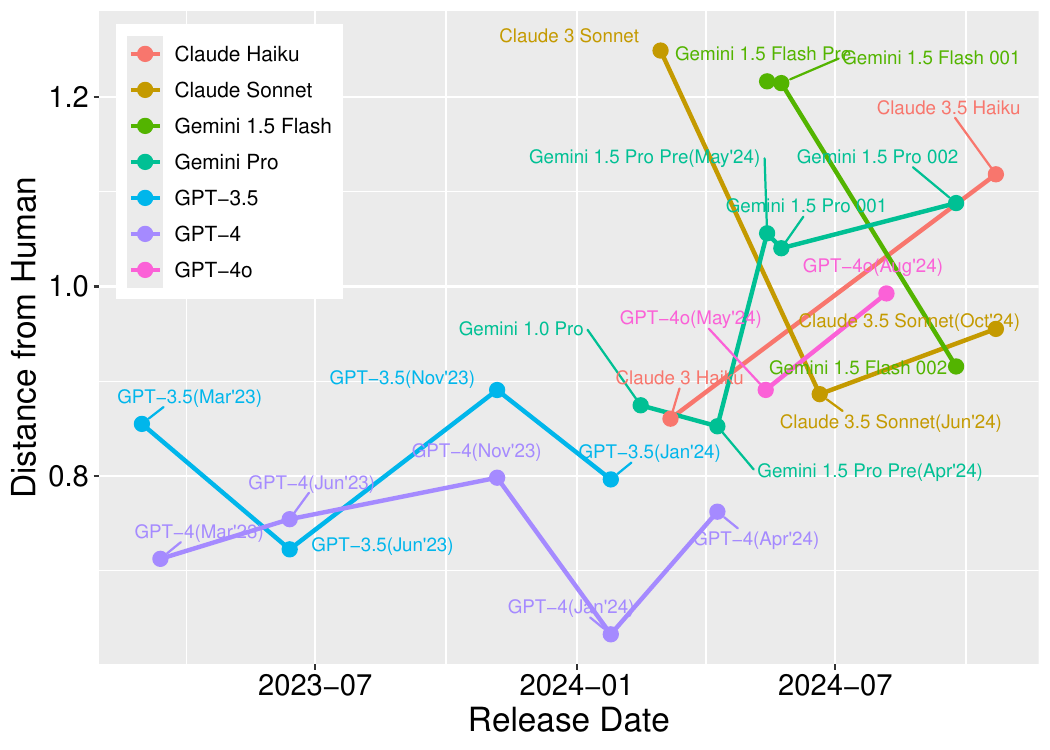} % Here is how to import EPS art
\end{center}
%\end{adjustwidth}
\caption{{\bf Temporal changes in distance from human judgments across proprietary model families.}
Different model families are represented by different colors. Model names are labeled.}
\label{fig:distance_trends_by_group}
\end{figure}

Proprietary models showed more stable but still variable patterns across updates (Fig \ref{fig:distance_trends_by_group}).
The GPT-4 family maintained relatively consistent alignment (distances = 0.7--0.8) from March 2023 to January 2024, with the January version achieving the best alignment (distance: 0.6) before slightly increasing in April (distance: 0.8).
GPT-3.5 showed similar stability (distances = 0.7--0.9), while the newer GPT-4o family demonstrated slightly larger distances (0.9--1.0) compared to traditional GPT-4.

The Claude and Gemini families exhibited more pronounced update effects.
Claude Sonnet's update from 3 to 3.5 improved alignment (distance decreased from 1.2 to 0.9), but the October 2024 update slightly increased divergence (distance: 1.0). 
The Gemini family maintained relatively stable alignment through version 1.5 Pro Preview (distance: 0.9), but later versions showed increased distances (1.1), with Flash models consistently showing larger divergence than Pro models.

These patterns suggest that model updates, while often aimed at improving general capabilities, do not guarantee better alignment with human moral preferences. This observation raises important questions about the relationship between general model improvements and ethical reasoning capabilities.

\subsection*{Comprehensive analysis of moral judgments}
\begin{figure}[!h]
%\begin{adjustwidth}{-2.25in}{0in} 
\begin{center}
\includegraphics[width=130mm]{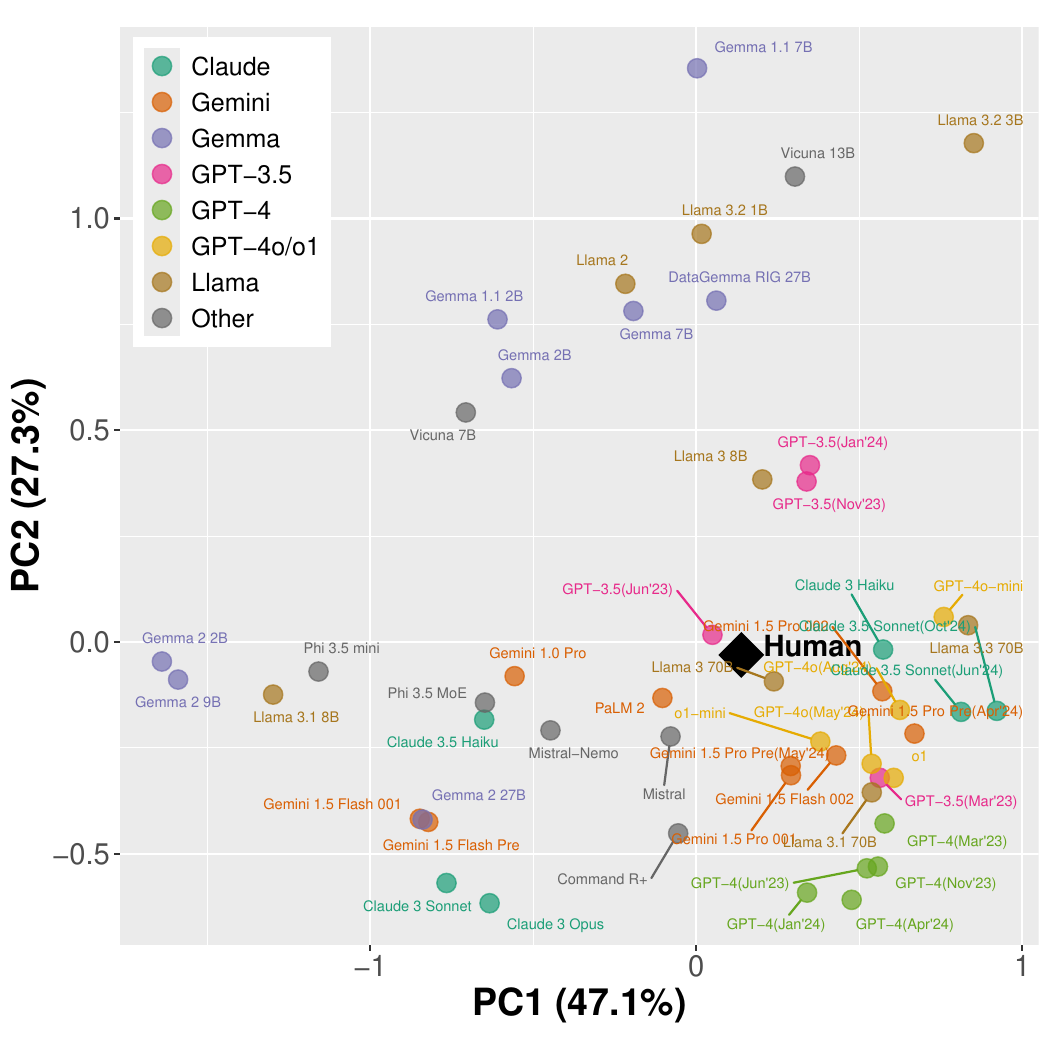} % Here is how to import EPS art
\end{center}
%\end{adjustwidth}
\caption{{\bf Principal component analysis of LLM moral judgments compared to human preferences.}
the first two principal components derived from AMCE values across nine moral preference categories. Different model families are represented by different colors, and human judgment is marked with a black diamond.}
\label{fig:pca_model_preferences}
\end{figure}

To synthesize our findings and visualize the overall landscape of moral judgment patterns, we conducted PCA using AMCE values across all nine ethical categories (Fig \ref{fig:pca_model_preferences}).
This analysis captured 74.2\% of the total variance in the first two principal components (PC1: 46.1\%, PC2: 28.1\%), providing a comprehensive two-dimensional representation of moral judgment similarities.

This analysis revealed several characteristic patterns.
First, the GPT-4 family, GPT-4o/o1 family, Claude 3 Haiku, Claude 3.5 Sonnet, Gemini Pro 1.5 family, and GPT-3.5 family formed clusters relatively close to human judgments.
Notably, these models concentrated in the positive region of the first principal component, suggesting commonalities in their judgment patterns.

Conversely, Claude 3 Sonnet and Opus, along with Gemini 1.5 Flash Pre and Flash 001, positioned in strongly negative regions of the first principal component, demonstrating judgment patterns distinct from humans and other proprietary models. 
The Gemma 2 family (2B, 9B) showed large positive values in the second principal component, confirming distinctively different judgment patterns from other models.

Interestingly, these results aligned with our previous distance analysis in multiple aspects.
Models showing smaller distances from human judgments tended to position closer to humans in this two-dimensional plane.
Similar trends were observed regarding model size effects.
For instance, within the Llama family, 70B models (Llama 3, 3.1, and 3.3) positioned near humans, while smaller 1-8B models distributed farther away.

Changes due to model updates were also clearly observable in this two-dimensional plane.
For example, GPT-3.5 approached human positions most closely in the June 2023 version but moved away with subsequent updates.
Similarly, while Claude Sonnet moved closer to humans with the update from 3 to 3.5, it slightly diverged with the October version of 3.5.
These trajectories support our previous finding that updates do not necessarily lead to closer alignment with human judgments.

\section*{Discussion}
In this study, we investigated two critical questions in AI ethics: how different LLM architectures approach moral decision-making in autonomous driving contexts, and how these moral judgments evolve through model updates.
By expanding our analysis to 52 different LLMs using the Moral Machine experimental framework (Figs \ref{fig:llm_preference_radar_chart} and \ref{fig:distance_wrt_modeltype}), we revealed several key patterns.
Proprietary models and open-source models exceeding 10B parameters demonstrated relatively small distances from human judgments (Figs \ref{fig:distance_wrt_opensource} and \ref{fig:modelparam_vs_distance}), showing consistent alignment with human preferences in fundamental moral decisions, such as prioritizing humans over animals and favoring the preservation of more lives.
However, model updates did not necessarily lead to closer alignment with human judgments, and variations in judgment tendencies were observed across updates (Fig \ref{fig:distance_trends_by_group}).
This comprehensive analysis provides crucial insights into both the current state and evolution of moral reasoning capabilities in large language models.

Our analysis revealed a complex relationship between model characteristics and moral judgment capabilities. While the incomplete alignment between LLM and human moral judgments was expected, as these models are not explicitly trained for moral reasoning, we identified several significant patterns that inform our understanding of how different architectures approach ethical decision-making.

A key finding was the significant negative correlation between model size and distance from human judgments in open-source models (Fig \ref{fig:modelparam_vs_distance}). 
This relationship aligns with general scaling laws in language model behavior, suggesting that increased model capacity might naturally lead to more human-like moral reasoning.
However, our cross-version analysis revealed that this relationship is not straightforward: within the same model family, size increases did not always result in improved alignment with human judgments, indicating that factors such as architecture design and training methodology play crucial roles.

These findings have important implications for the practical implementation of AI ethics in autonomous driving systems. The GPT-4 family achieved the closest alignment with human judgments (distances = 0.6--0.8), while among open-source alternatives, Llama 70B demonstrated promising performance (distances = 0.7–0.8).
However, real-world deployment faces a crucial challenge: the trade-off between ethical reasoning quality and computational efficiency.
Smaller models, which are more practical for real-time decision-making, showed notably larger divergence from human judgments (distances = 1.2--1.5).
This creates a significant challenge for implementing ethical AI in autonomous systems where both moral reliability and operational efficiency are essential.

Among these implementation challenges, the computational efficiency requirements are particularly critical in autonomous driving, where decisions must be made in milliseconds. 
While our study demonstrates that larger models generally make better ethical decisions, their computational requirements make them impractical for real-time deployment.
This challenge necessitates innovative approaches such as model distillation \cite{tang2023domain} or specialized ethical reasoning modules that can maintain high-quality moral judgment while meeting the strict latency requirements of autonomous systems.

The identified trade-off between model size and ethical reasoning quality thus presents fundamental challenges for practical implementation.
While recent research has produced LLMs specifically fine-tuned for autonomous driving tasks \cite{yang2023llm4drive}, these efforts have primarily focused on technical capabilities (perception, control, navigation) rather than ethical reasoning.
Our findings suggest that developing practically viable AI systems requires addressing three key challenges: optimizing smaller models for better ethical reasoning, managing excessive ethical preferences, and maintaining cultural adaptability.

The tendency of LLMs to exhibit exaggerated ethical preferences represents a particular concern.
Our analysis showed that while models often directionally aligned with human preferences (e.g., prioritizing humans over pets, pedestrians over passengers), they frequently demonstrated substantially stronger preferences than humans.
While such strong safety preferences partially align with established guidelines, as exemplified by the German Ethics Code on Automated and Connected Driving \cite{luetge2017german}, their excessive nature could lead to suboptimal decision-making in complex real-world scenarios.
Specifically, while Guidelines 5 and 7 emphasize protecting vulnerable road users and preventing personal injury, the observed extreme preferences in LLMs suggest a need for more nuanced ethical calibration.

Our analysis revealed particularly significant findings regarding utilitarian decision-making in LLMs.
The models showed extreme emphasis on maximizing lives saved, reflecting a strong utilitarian bias that exceeds typical human preferences.
This bias likely stems from the predominance of Western individualistic and utilitarian values in training data \cite{awad2018moral,kagitcibasi1997individualism}, raising important questions about cultural representation in AI ethics.
This observation becomes especially significant when considered against formal ethical frameworks like the German Ethics Code, whose Guideline 2 explicitly warns against purely utilitarian approaches in autonomous vehicle decision-making.

The analysis also uncovered complex patterns in how LLMs handle social biases and protective principles.
We observed inverse patterns to human judgments in cases involving physical fitness and social status, where LLMs showed stronger preferences for protecting potentially vulnerable groups.
Similarly, while LLMs aligned with human tendencies in gender-related decisions and legal compliance, they demonstrated significantly amplified preferences.
These patterns suggest that LLMs may be extracting and amplifying certain ethical principles from their training data, such as historical protective principles (e.g., `Women and Children First'' principle \cite{annas1995women}) and fundamental concepts of justice \cite{sandel2010justice}, but applying them with excessive rigidity.

Our findings highlight fundamental challenges for implementing ethical AI in autonomous driving systems. A critical issue emerged from the contrast between human moral flexibility and the rigid, often amplified moral preferences displayed by LLMs. This rigidity becomes particularly problematic when considered alongside recent findings \cite{vida2024decoding} demonstrating that LLMs exhibit varying moral biases across different languages, suggesting inconsistent ethical reasoning even within individual models.

These observations raise two critical concerns for practical implementation.
First, the potential for AI systems to influence human decision-making \cite{krugel2023chatgpt} means that LLMs' amplified moral preferences could disproportionately shape social attitudes toward ethical decisions in autonomous driving.
Second, the observed variation in moral judgments across languages and cultures challenges the goal of developing globally deployable autonomous systems.
These issues are particularly significant when considered alongside established ethical guidelines. For instance, the German Ethics Code's Guideline 9 emphasizes equal treatment regardless of demographic factors, but this principle appears challenging to implement given the observed biases in LLM decision-making.

Another critical consideration for practical deployment is the need for transparency and explainability in LLMs' ethical decision-making processes.
While our study quantifies the alignment between LLM and human moral preferences, the black-box nature of these models poses significant challenges for real-world implementation, particularly in life-critical decisions.
Unlike rule-based systems, LLMs' ethical judgments emerge from complex neural network interactions, making them difficult to interpret or verify.
This lack of transparency is particularly problematic in autonomous driving scenarios where stakeholders need to understand and trust the basis for ethical decisions \cite{atakishiyev2024explainable}.

Our findings raise significant concerns about the cultural adaptability of AI ethics in autonomous systems.
The observed tendency of LLMs to reflect predominantly Western values and apply ethical principles in extreme ways could significantly impact the global acceptance of autonomous driving systems.
This challenge is compounded by our observation that LLMs' moral judgments can diverge substantially from human preferences across different cultural and linguistic contexts, suggesting a need for more culturally adaptive approaches to ethical AI implementation.

These cultural adaptability challenges point to specific implementation considerations for autonomous driving systems. Regions with distinct driving cultures and ethical priorities may require different thresholds for risk assessment and decision-making parameters. This is particularly crucial in areas where different cultural zones intersect, such as international borders or multicultural urban centers, where autonomous vehicles must navigate varying expectations of driving behavior while maintaining consistent safety standards.

A key limitation of our study stems from our use of globally aggregated moral preferences as the benchmark for human judgments.
Moral decisions, particularly those involving life-and-death choices, are deeply influenced by cultural and social contexts. For example, the strong utilitarian preferences we observed in LLMs might align with some cultural perspectives while conflicting with others. 
While our analysis demonstrates clear patterns in LLM moral reasoning, it does not fully capture the nuanced variations in human moral preferences across different cultural contexts.

Our methodology, based on the Moral Machine experiment framework, carries inherent limitations that affect the interpretation of our findings.
While this framework enabled systematic comparison across models, it simplifies the complexity of real-world ethical decisions through binary choices and trolley-type dilemmas.
As noted in \cite{takemoto2024moral}, this approach cannot capture nuanced factors such as road users' intentions or evaluate everyday moral decisions in low-risk traffic situations. Recent work has proposed more ecologically valid approaches, such as combining virtual reality-based traffic scenarios with psychological frameworks like the Agent-Deed-Consequences model \cite{cecchini2023moral}.

The emergence of multimodal large language models (MLLMs) offers potential pathways for addressing these methodological limitations.
Recent research has demonstrated MLLMs' capability to integrate visual information with language understanding in autonomous driving contexts \cite{cui2024survey,zhou2024vision}, suggesting possibilities for more comprehensive ethical evaluation frameworks.
Unlike the text-based scenarios used in our study, MLLMs could enable assessment of ethical decision-making in more realistic, visually-rendered traffic scenarios, potentially offering deeper insights into AI moral reasoning in real-world contexts.

Despite these limitations, our comprehensive analysis of 52 different LLMs provides three key contributions to understanding AI ethics in autonomous systems.
First, we established a clear relationship between model size and ethical judgment capabilities, offering insights for future model development.
Second, our systematic characterization of judgment patterns across model families revealed both consistent trends and concerning biases in AI moral reasoning.
Third, we identified specific implementation challenges that must be addressed for practical deployment of ethical AI in autonomous vehicles. These findings, combined with our assessment of current methodological constraints, provide crucial guidance for both the technical development and ethical design of future autonomous driving systems.

\section*{Supporting information}

% Include only the SI item label in the paragraph heading. Use the \nameref{label} command to cite SI items in the text.
\paragraph*{S1 Fig.}
\label{S1_Fig}
{\bf Differences in AMCE values between LLMs and human preferences.} Heatmap showing the differences between LLM and human AMCE values for nine moral preference categories.

\paragraph*{S1 Table.}
\label{S1_Table}
{\bf AMCE values for nine moral preference categories across all analyzed LLMs.}

\paragraph*{S2 Table.}
\label{S2_Table}
{\bf Comprehensive model information and evaluation metrics.}
Summary of all analyzed LLMs, including their characteristics and performance metrics. The `open-source' column indicates model accessibility (yes/no), `Size' shows the number of parameters in billions where available, `Date' indicates the model version's release date, `Distance' shows the Euclidean distance between the model's and human AMCE values across nine moral preference categories, and `valid response rate' represents the proportion of valid responses in the evaluation scenarios.

% \section*{Acknowledgments}

\nolinenumbers

% Either type in your references using
% \begin{thebibliography}{}
% \bibitem{}
% Text
% \end{thebibliography}
%
% or
%
% Compile your BiBTeX database using our plos2015.bst
% style file and paste the contents of your .bbl file
% here. See http://journals.plos.org/plosone/s/latex for 
% step-by-step instructions.
% 
% \begin{thebibliography}{10}

% \bibitem{bib1}
% Conant GC, Wolfe KH.
% \newblock {{T}urning a hobby into a job: how duplicated genes find new
%   functions}.
% \newblock Nat Rev Genet. 2008 Dec;9(12):938--950.

% \bibitem{bib2}
% Ohno S.
% \newblock Evolution by gene duplication.
% \newblock London: George Alien \& Unwin Ltd. Berlin, Heidelberg and New York:
%   Springer-Verlag.; 1970.

% \bibitem{bib3}
% Magwire MM, Bayer F, Webster CL, Cao C, Jiggins FM.
% \newblock {{S}uccessive increases in the resistance of {D}rosophila to viral
%   infection through a transposon insertion followed by a {D}uplication}.
% \newblock PLoS Genet. 2011 Oct;7(10):e1002337.

% \end{thebibliography}

\bibliography{references}

\begin{thebibliography}{10}

\bibitem{fraiwan2023review}
Fraiwan M, Khasawneh N.
\newblock A review of chatgpt applications in education, marketing, software engineering, and healthcare: Benefits, drawbacks, and research directions.
\newblock arXiv preprint arXiv:230500237. 2023;.

\bibitem{sallam2023chatgpt}
Sallam M.
\newblock ChatGPT utility in healthcare education, research, and practice: systematic review on the promising perspectives and valid concerns.
\newblock In: Healthcare. vol.~11. MDPI; 2023. p. 887.

\bibitem{ray2023chatgpt}
Ray PP.
\newblock ChatGPT: A comprehensive review on background, applications, key challenges, bias, ethics, limitations and future scope.
\newblock Internet of Things and Cyber-Physical Systems. 2023;3:121--154.

\bibitem{chowdhury2024harnessing}
Chowdhury AK, Sujon SR, Shafi MSS, Ahmmad T, Ahmed S, Hasib KM, et~al.
\newblock Harnessing large language models over transformer models for detecting Bengali depressive social media text: A comprehensive study.
\newblock Natural Language Processing Journal. 2024;7:100075.

\bibitem{chatgpt}
OpenAI. Introducing ChatGPT; 2022.
\newblock OpenAI.
\newblock Available from: \url{https://openai.com/index/chatgpt/}.

\bibitem{achiam2023gpt}
Achiam J, Adler S, Agarwal S, Ahmad L, Akkaya I, Aleman FL, et~al.
\newblock GPT-4 technical report.
\newblock arXiv preprint arXiv:230308774. 2023;.

\bibitem{bostrom2018ethics}
Bostrom N, Yudkowsky E.
\newblock The ethics of artificial intelligence.
\newblock In: Artificial intelligence safety and security. Chapman and Hall/CRC; 2018. p. 57--69.

\bibitem{nath2020problem}
Nath R, Sahu V.
\newblock The problem of machine ethics in artificial intelligence.
\newblock AI \& society. 2020;35(1):103--111.

\bibitem{hagendorff2020ethics}
Hagendorff T.
\newblock The ethics of AI ethics: An evaluation of guidelines.
\newblock Minds and machines. 2020;30(1):99--120.

\bibitem{bonnefon2016social}
Bonnefon JF, Shariff A, Rahwan I.
\newblock The social dilemma of autonomous vehicles.
\newblock Science. 2016;352(6293):1573--1576.

\bibitem{awad2018moral}
Awad E, Dsouza S, Kim R, Schulz J, Henrich J, Shariff A, et~al.
\newblock The moral machine experiment.
\newblock Nature. 2018;563(7729):59--64.

\bibitem{faulhaber2019human}
Faulhaber AK, Dittmer A, Blind F, W{\"a}chter MA, Timm S, S{\"u}tfeld LR, et~al.
\newblock Human decisions in moral dilemmas are largely described by utilitarianism: Virtual car driving study provides guidelines for autonomous driving vehicles.
\newblock Science and engineering ethics. 2019;25:399--418.

\bibitem{gill2021ethical}
Gill T.
\newblock Ethical dilemmas are really important to potential adopters of autonomous vehicles.
\newblock Ethics and Information Technology. 2021;23(4):657--673.

\bibitem{chen2023feedback}
Chen H, Yuan K, Huang Y, Guo L, Wang Y, Chen J.
\newblock Feedback is all you need: from ChatGPT to autonomous driving.
\newblock Science China Information Sciences. 2023;66(6):1--3.

\bibitem{gao2023chat}
Gao Y, Tong W, Wu EQ, Chen W, Zhu G, Wang FY.
\newblock Chat with ChatGPT on interactive engines for intelligent driving.
\newblock IEEE Transactions on Intelligent Vehicles. 2023;8(3):2034--2036.

\bibitem{du2023chat}
Du H, Teng S, Chen H, Ma J, Wang X, Gou C, et~al.
\newblock Chat with chatgpt on intelligent vehicles: An ieee tiv perspective.
\newblock IEEE Transactions on Intelligent Vehicles. 2023;8(3):2020--2026.

\bibitem{lei2023chatgpt}
Lei L, Zhang H, Yang SX.
\newblock ChatGPT in connected and autonomous vehicles: benefits and challenges.
\newblock Intell Robot. 2023;3(2):145--148.

\bibitem{yang2023llm4drive}
Yang Z, Jia X, Li H, Yan J.
\newblock Llm4drive: A survey of large language models for autonomous driving.
\newblock arXiv e-prints. 2023; p. arXiv--2311.

\bibitem{cui2024survey}
Cui C, Ma Y, Cao X, Ye W, Zhou Y, Liang K, et~al.
\newblock A survey on multimodal large language models for autonomous driving.
\newblock In: Proceedings of the IEEE/CVF Winter Conference on Applications of Computer Vision; 2024. p. 958--979.

\bibitem{zhou2024vision}
Zhou X, Liu M, Yurtsever E, Zagar BL, Zimmer W, Cao H, et~al.
\newblock Vision language models in autonomous driving: A survey and outlook.
\newblock IEEE Transactions on Intelligent Vehicles. 2024;.

\bibitem{dewitt2019moral}
Dewitt B, Fischhoff B, Sahlin NE.
\newblock 'Moral machine'experiment is no basis for policymaking.
\newblock Nature. 2019;567(7746):31--31.

\bibitem{bigman2020life}
Bigman YE, Gray K.
\newblock Life and death decisions of autonomous vehicles.
\newblock Nature. 2020;579(7797):E1--E2.

\bibitem{etienne2021dark}
Etienne H.
\newblock The dark side of the ‘Moral Machine’and the fallacy of computational ethical decision-making for autonomous vehicles.
\newblock Law, Innovation and Technology. 2021;13(1):85--107.

\bibitem{furey2021s}
Furey H, Hill S.
\newblock MIT’s moral machine project is a psychological roadblock to self-driving cars.
\newblock AI and Ethics. 2021;1(2):151--155.

\bibitem{lacroix2022moral}
LaCroix T.
\newblock Moral dilemmas for moral machines.
\newblock AI and Ethics. 2022;2(4):737--746.

\bibitem{schuessler2024probability}
Schuessler D.
\newblock The probability problems of the Moral Machine Experiment.
\newblock AI and Ethics. 2024;4(2):501--510.

\bibitem{winfield2019machine}
Winfield AF, Michael K, Pitt J, Evers V.
\newblock Machine ethics: The design and governance of ethical AI and autonomous systems [scanning the issue].
\newblock Proceedings of the IEEE. 2019;107(3):509--517.

\bibitem{etienne2022ai}
Etienne H.
\newblock When AI ethics goes astray: A case study of autonomous vehicles.
\newblock Social science computer review. 2022;40(1):236--246.

\bibitem{meyer2022baby}
Meyer-Waarden L, Cloarec J.
\newblock “Baby, you can drive my car”: Psychological antecedents that drive consumers’ adoption of AI-powered autonomous vehicles.
\newblock Technovation. 2022;109:102348.

\bibitem{atakishiyev2024explainable}
Atakishiyev S, Salameh M, Yao H, Goebel R.
\newblock Explainable artificial intelligence for autonomous driving: A comprehensive overview and field guide for future research directions.
\newblock IEEE Access. 2024;.

\bibitem{krugel2023chatgpt}
Kr{\"u}gel S, Ostermaier A, Uhl M.
\newblock ChatGPT’s inconsistent moral advice influences users’ judgment.
\newblock Scientific Reports. 2023;13(1):4569.

\bibitem{jin2024multilingual}
Jin Z, Levine S, Kleiman-Weiner M, Piatti G, Liu J, Adauto FG, et~al.
\newblock Multilingual Trolley Problems for Language Models.
\newblock arXiv preprint arXiv:240702273. 2024;.

\bibitem{bruers2014review}
Bruers S, Braeckman J.
\newblock A review and systematization of the trolley problem.
\newblock Philosophia. 2014;42:251--269.

\bibitem{takemoto2024moral}
Takemoto K.
\newblock The moral machine experiment on large language models.
\newblock Royal Society open science. 2024;11(2):231393.

\bibitem{anthropic2024claude}
Anthropic A.
\newblock The claude 3 model family: Opus, sonnet, haiku.
\newblock Claude-3 Model Card. 2024;1.

\bibitem{team2023gemini}
Team G, Anil R, Borgeaud S, Wu Y, Alayrac JB, Yu J, et~al.
\newblock Gemini: a family of highly capable multimodal models.
\newblock arXiv preprint arXiv:231211805. 2023;.

\bibitem{dubey2024llama}
Dubey A, Jauhri A, Pandey A, Kadian A, Al-Dahle A, Letman A, et~al.
\newblock The llama 3 herd of models.
\newblock arXiv preprint arXiv:240721783. 2024;.

\bibitem{team2024gemma}
Team G, Mesnard T, Hardin C, Dadashi R, Bhupatiraju S, Pathak S, et~al.
\newblock Gemma: Open models based on gemini research and technology.
\newblock arXiv preprint arXiv:240308295. 2024;.

\bibitem{GPT4o}
OpenAI. Hello GPT-4o; 2024.
\newblock OpenAI.
\newblock Available from: \url{https://openai.com/index/hello-gpt-4o/}.

\bibitem{OpenAIo1}
OpenAI. Introducing OpenAI o1; 2024.
\newblock OpenAI.
\newblock Available from: \url{https://openai.com/o1/}.

\bibitem{reid2024gemini}
Reid M, Savinov N, Teplyashin D, Lepikhin D, Lillicrap T, Alayrac Jb, et~al.
\newblock Gemini 1.5: Unlocking multimodal understanding across millions of tokens of context.
\newblock arXiv preprint arXiv:240305530. 2024;.

\bibitem{anthropic2024claude3.5}
Anthropic A.
\newblock Claude 3.5 sonnet model card addendum.
\newblock Claude-35 Model Card. 2024;.

\bibitem{team2024gemma2}
Team G, Riviere M, Pathak S, Sessa PG, Hardin C, Bhupatiraju S, et~al.
\newblock Gemma 2: Improving open language models at a practical size.
\newblock arXiv preprint arXiv:240800118. 2024;.

\bibitem{radhakrishnan2024knowing}
Radhakrishnan P, Chen J, Xu B, Ramaswami P, Pho H, Olmos A, et~al.
\newblock Knowing When to Ask--Bridging Large Language Models and Data.
\newblock arXiv preprint arXiv:240913741. 2024;.

\bibitem{vicuna2023}
Chiang WL, Li Z, Lin Z, Sheng Y, Wu Z, Zhang H, et~al.. Vicuna: An Open-Source Chatbot Impressing GPT-4 with 90\%* ChatGPT Quality; 2023.
\newblock Available from: \url{https://lmsys.org/blog/2023-03-30-vicuna/}.

\bibitem{jiang2023mistral}
Jiang AQ, Sablayrolles A, Mensch A, Bamford C, Chaplot DS, Casas Ddl, et~al.
\newblock Mistral 7B.
\newblock arXiv preprint arXiv:231006825. 2023;.

\bibitem{MistralNeMo}
MistralAIteam. Mistral NeMo; 2024.
\newblock Mistral AI.
\newblock Available from: \url{https://mistral.ai/news/mistral-nemo/}.

\bibitem{CommandRplus2024}
Gomez A. Introducing Command R+: A Scalable LLM Built for Business; 2024.
\newblock The Cohere Blog.
\newblock Available from: \url{https://cohere.com/blog/command-r-plus-microsoft-azure}.

\bibitem{abdin2024phi}
Abdin M, Jacobs SA, Awan AA, Aneja J, Awadallah A, Awadalla H, et~al.
\newblock Phi-3 technical report: A highly capable language model locally on your phone.
\newblock arXiv preprint arXiv:240414219. 2024;.

\bibitem{hainmueller2014causal}
Hainmueller J, Hopkins DJ, Yamamoto T.
\newblock Causal inference in conjoint analysis: Understanding multidimensional choices via stated preference experiments.
\newblock Political analysis. 2014;22(1):1--30.

\bibitem{zhang2024closing}
Zhang G, Jin Q, Zhou Y, Wang S, Idnay B, Luo Y, et~al.
\newblock Closing the gap between open source and commercial large language models for medical evidence summarization.
\newblock npj Digital Medicine. 2024;7(1):239.

\bibitem{tang2023domain}
Tang Y, Da~Costa AAB, Zhang X, Patrick I, Khastgir S, Jennings P.
\newblock Domain knowledge distillation from large language model: An empirical study in the autonomous driving domain.
\newblock In: 2023 IEEE 26th International Conference on Intelligent Transportation Systems (ITSC). IEEE; 2023. p. 3893--3900.

\bibitem{luetge2017german}
Luetge C.
\newblock The German ethics code for automated and connected driving.
\newblock Philosophy \& Technology. 2017;30:547--558.

\bibitem{kagitcibasi1997individualism}
Kagitcibasi C.
\newblock Individualism and collectivism.
\newblock Handbook of cross-cultural psychology. 1997;3:1--49.

\bibitem{annas1995women}
Annas GJ.
\newblock Women and children first.
\newblock New England Journal of Medicine. 1995;333:1647.

\bibitem{sandel2010justice}
Sandel MJ.
\newblock Justice: What's the Right Thing to Do?
\newblock Farrar, Straus and Giroux; 2010.
\newblock Available from: \url{https://books.google.co.jp/books?id=BrdNDG7TTUEC}.

\bibitem{vida2024decoding}
Vida K, Damken F, Lauscher A.
\newblock Decoding Multilingual Moral Preferences: Unveiling LLM's Biases Through the Moral Machine Experiment.
\newblock In: Proceedings of the AAAI/ACM Conference on AI, Ethics, and Society. vol.~7; 2024. p. 1490--1501.

\bibitem{cecchini2023moral}
Cecchini D, Brantley S, Dubljevi{\'c} V.
\newblock Moral judgment in realistic traffic scenarios: moving beyond the trolley paradigm for ethics of autonomous vehicles.
\newblock AI \& SOCIETY. 2023; p. 1--12.

\end{thebibliography}

\end{document}